\documentclass[11pt, a4paper, reqno]{article}

\usepackage[utf8]{inputenc}
\usepackage[T1]{fontenc}
\usepackage{lmodern}
\usepackage{amsmath, amssymb, amsthm, amsfonts}
\usepackage{mathtools}
\usepackage{mathrsfs}
\usepackage{geometry}
\usepackage{xcolor}
\usepackage{enumitem}
\usepackage{graphicx}
\usepackage{authblk}
\usepackage[colorlinks=true, linkcolor=blue, citecolor=red, urlcolor=blue]{hyperref}

\theoremstyle{plain}
\newtheorem{theorem}{Theorem}[section]

\theoremstyle{definition}

\theoremstyle{remark}

\numberwithin{equation}{section}

\title{\textbf{A Rigorous Proof of Metric Blow-Up for Pseudospherical Metrics Associated with the Degasperis--Procesi Equation}}

\author{ Zhenhua Shi$^{1,2}$, Mingyue Guo$^{1}$
	\vspace{4mm}\\
	$^{1}$\small{School of Mathematics, Northwest University, Xi'an 710069,  P.R. China}\\
	$^{2}$\small {Center for Nonlinear Studies, Northwest University, Xi'an 710069, P.R. China}\\
	\small {Northwest University, Xi'an 710069, P.R. China}}

\date{}

\begin{document}
	
	\maketitle
	
	\begin{abstract}
		\noindent
		This paper studies finite-time blow-up of pseudospherical metrics induced by solutions of the Cauchy problem for the Degasperis--Procesi equation. 
		For initial momentum profiles satisfying suitable left--right sign conditions, we use the method of characteristics, the momentum-transport formula, the Green's-function representation, and Riccati-type differential inequalities to analyze the metric along a distinguished characteristic. 
		We prove that the associated coframe remains non-degenerate before the critical time, so that the induced pseudospherical metric is well defined in the precritical region. 
		Moreover, as wave breaking is approached, the metric component \(g_{22}\) diverges to \(+\infty\); when \(\mu\neq0\), the mixed component \(g_{12}\) also blows up in absolute value. 
		Thus, finite-time wave breaking for the Degasperis--Procesi equation is shown to induce blow-up of certain components of the corresponding pseudospherical metric.
		
		\vspace{1em}
	\end{abstract}
	
	\tableofcontents
	
	\section{Introduction}
	\label{sec:intro}
	In the study of nonlinear shallow water theory and geometrically integrable systems, 
	the \(b\)-family constitutes an important class of nonlinear shallow water models. The general form of this family is
	\[
	u_{t}-u_{txx}+(b+1)uu_{x}=bu_{x}u_{xx}+uu_{xxx},
	\]
	where $b\in\mathbb{R}$ is a constant parameter. This family plays an important role in describing nonlinear propagation of shallow water waves, dispersive effects, and wave breaking. Related developments in the theory of nonlinear shallow water equations include global existence versus finite-time wave breaking, geometric descriptions of breaking waves via weak Riemannian structures on diffeomorphism groups, and shock-wave formation for the Degasperis--Procesi~(DP) equation~\cite{Constantin2000,ConstantinEscher1998a,ConstantinEscher1998b,Lundmark2007}. In particular, among the various choices of the parameter, two members occupy a central place in the theory of complete integrability: when $b=2$, one obtains the Camassa--Holm (CH) equation, while when $b=3$, one obtains the DP equation. Both equations possess peakon solutions, rich nonlinear dynamical structures, and finite-time wave breaking; nevertheless, they differ significantly in their Lax pairs, zero-curvature representations, and associated geometric structures~\cite{Camassa1993,Degasperis1999,Degasperis2002,Escher2008}.
	
	The CH equation is closely related to an $\mathfrak{sl}(2,\mathbb{R})$-valued zero curvature representation, and therefore it can be incorporated rather naturally into the Chern--Tenenblat theory of pseudospherical equations \cite{Chern1986,Reyes2002,Sasaki1979}. By contrast, the DP equation is usually associated with an $\mathfrak{sl}(3,\mathbb{R})$-valued zero curvature representation, so its connection with two-dimensional pseudospherical geometry is not immediately evident. However, previous studies have shown that the DP equation can still be described in terms of a suitable triple of one-forms defining pseudospherical surfaces~\cite{CastroSilva2015,CastroSilva2016}. In other words, solutions of the DP equation induce differential forms satisfying the Maurer--Cartan structure equations, thereby determining a two-dimensional Riemannian metric of constant negative Gaussian curvature.
	
	Let $u=u(x,t)$ be a solution of the DP equation, and set
	\[
	m=u-u_{xx}.
	\]
	For sufficiently regular solutions, and with \(m=u-u_{xx}\), the DP equation admits two equivalent formulations: the nonlocal evolution form
	\[
	u_t+u u_x+\frac{3}{2}\partial_x(1-\partial_x^2)^{-1}(u^2)=0,
	\]
	and the momentum-transport form
	\[
	m_t+u m_x+3u_xm=0.
	\]
	These formulations play a central role in the study of the well-posedness of the Cauchy problem, characteristic-line evolution, and the mechanism of wave breaking for the DP equation~\cite{Yin2003,Yin2004,Liu2006,Liu2007,Escher2008,Henry2008}. In particular, when the initial momentum satisfies certain sign conditions, the solution may undergo wave breaking in finite time: the solution remains bounded, while the spatial derivative $u_{x}$ becomes unbounded in finite time.
	
	On the other hand, the pseudospherical structure induced by a solution of the DP equation can be described by a triple of one-forms of the form
	\[
	\omega_{i}=f_{i1}dx+f_{i2}dt, \quad i=1,2,3.
	\]
	These one-forms satisfy the structure equations of pseudospherical surfaces,
	\[
	d\omega_{1}=\omega_{3}\wedge\omega_{2}, \quad d\omega_{2}=\omega_{1}\wedge\omega_{3}, \quad d\omega_{3}=\omega_{1}\wedge\omega_{2}.
	\]
	When $\omega_{1}\wedge\omega_{2} \neq 0$, the forms $\omega_{1}$ and $\omega_{2}$ constitute a local coframe and hence define the first fundamental form
	\[
	g=\omega_{1}^{2}+\omega_{2}^{2}.
	\]
	Thus a solution of the DP equation is not only an evolving function in the sense of partial differential equations, but also geometrically determines a pseudospherical metric.
	
	This raises a natural question: what happens to the pseudospherical metric induced by a solution of the DP equation when the solution undergoes wave breaking? More specifically, does the wave-breaking singularity of the PDE solution manifest itself in the induced first fundamental form, leading to finite-time blow-up of certain metric components? This question connects the singularity theory of nonlinear partial differential equations with the differential geometry of pseudospherical surfaces, and it is an important step toward understanding the geometric structure of the DP equation.
	
	This paper investigates that question and gives a rigorous proof of metric blow-up for the pseudospherical metric associated with the DP equation. Our proof combines the method of characteristics, the momentum-transport formulation, and the Green's-function convolution representation to derive Riccati-type differential 
	inequalities along characteristic curves. First, we prove that, along the relevant characteristic and on the associated precritical region, the induced coframe remains non-degenerate, so that the corresponding first fundamental form is well defined. Second, by deriving Riccati-type differential inequalities along this characteristic, we show that, as \(t\to T_0^-\), the metric component \(g_{22}(\gamma(t))\) diverges to \(+\infty\). Furthermore, when the geometric parameter satisfies \(\mu \neq 0\), the mixed component also blows up in the sense that \(|g_{12}(\gamma(t))|\to+\infty\). 
	In this way, we establish a rigorous connection between finite-time wave breaking for the DP equation and blow-up of its induced pseudospherical metric.
	
	The paper is organized as follows. Section 2 recalls the local, nonlocal, and momentum-transport formulations of the DP equation, as well as the induced pseudospherical one-forms and first fundamental form. Section 3 states the main theorem, namely the finite-time metric blow-up theorem. Section 4 gives a complete proof of the theorem. Section 5 explains the geometric significance of the result, especially its relation to coframe non-degeneracy and local isometric immersions. Section 6 concludes the paper.
	
	\section{Preliminaries and the Induced Pseudospherical Metric}
	\label{sec:math_preliminaries}
	In preparation for the proof of the metric blow-up theorem, this section first recalls several equivalent formulations of the DP equation, the relevant Sobolev regularity assumptions, and the pseudospherical one-forms and first fundamental form induced by the equation.
	
	\subsection{Equivalent Formulations and Regularity of the DP Equation}
	The DP equation is the special case of the $b$-family corresponding to $b=3$. Its local form is
	\[
	u_{t}-u_{txx}+4uu_{x}=3u_{x}u_{xx}+uu_{xxx}, \quad x\in\mathbb{R}, \quad t>0.
	\]
	This equation contains a mixed partial derivative term and a third-order spatial derivative term. In order to study its Cauchy problem and wave-breaking mechanism, it is convenient to introduce the one-dimensional Helmholtz operator
	\[
	\Lambda^2=1-\partial_x^2
	\]
	and its inverse
	\[
	\Lambda^{-2}=(1-\partial_x^2)^{-1}.
	\]
	On the real line, $\Lambda^{-2}$ can be represented as convolution with the Green's function,
	\[
	\Lambda^{-2}f=G*f, \quad G(x)=\frac{1}{2}e^{-|x|}.
	\]
	Consequently, the corresponding Cauchy problem can be written in the nonlocal form
	\[
	\begin{cases}
		u_t+u u_x+\dfrac{3}{2}\partial_x\Lambda^{-2}(u^2)=0,
		& x\in\mathbb R,\quad t>0,\\[0.4em]
		u(x,0)=u_0(x),
		& x\in\mathbb R.
	\end{cases}
	\]
	
	Define the momentum variable $m=u-u_{xx}$. Then the DP equation can also be written in momentum transport form as
	\[
	m_{t}+um_{x}+3u_{x}m=0.
	\]
	This form is very important for studying the evolution of characteristics and the propagation of the sign of the initial momentum. If \(q(t,\xi)\) denotes the characteristic flow generated by the velocity field \(u\), namely
	\[
	\frac{d}{dt}q(t,\xi)=u(q(t,\xi),t),\qquad q(0,\xi)=\xi,
	\]
	then the momentum-transport equation governs the evolution of \(m\) along characteristics:
	\[
	\frac{d}{dt}m(q(t,\xi),t)
	=
	-3u_x(q(t,\xi),t)m(q(t,\xi),t).
	\]
	
	Concerning well-posedness~\cite{Liu2006,Liu2007,Yin2003,Yin2004}, it is known that if 
	\(u_{0}\in H^{s}(\mathbb{R})\) with \(s>3/2\), 
	then there exists \(T>0\) such that the DP equation admits a unique local strong solution
	\[
	u\in C([0,T);H^{s}(\mathbb{R}))\cap C^{1}([0,T);H^{s-1}(\mathbb{R})).
	\]
	For the geometric construction below, we shall work with higher-regularity initial data, namely
	\[
	u_{0}\in H^{4}(\mathbb{R}).
	\]
	Then, on the maximal interval of existence \([0,T_{\max})\), the corresponding solution satisfies
	\[
	u\in C([0,T_{\max});H^{4}(\mathbb{R}))
	\cap C^{1}([0,T_{\max});H^{3}(\mathbb{R})).
	\]
	By the one-dimensional Sobolev embedding theorem, for every \(T'<T_{\max}\), 
	the functions \(u,u_x,u_{xx},u_{xxx}\) are continuous on 
	\(\mathbb{R}\times[0,T']\), which provides the classical regularity required for the subsequent geometric construction.
	
	\subsection{Pseudospherical One-Forms and the Coframe Construction}
	The Chern--Tenenblat theory of pseudospherical equations establishes a connection 
	between certain partial differential equations and surfaces of constant Gaussian curvature \(K=-1\) \cite{Chern1986,Kamran1995,Reyes2011}. 
	More precisely, suppose that each sufficiently regular solution of a partial differential equation 
	determines three one-forms
	\[
	\omega_i=f_{i1}\,dx+f_{i2}\,dt,\qquad i=1,2,3,
	\]
	which satisfy the structure equations
	\[
	d\omega_1=\omega_3\wedge\omega_2,\qquad 
	d\omega_2=\omega_1\wedge\omega_3,\qquad 
	d\omega_3=\omega_1\wedge\omega_2.
	\]
	If, in addition, \(\omega_1\wedge\omega_2\) is nowhere zero on a region, then 
	\(\omega_1\) and \(\omega_2\) form a local coframe there, and the first fundamental form
	\[
	g=\omega_1^2+\omega_2^2
	\]
	defines a metric of Gaussian curvature \(K=-1\). In this sense, the equation is said to describe 
	pseudospherical surfaces.
	
	For the DP equation, set
	\[
	m=u-u_{xx}, \quad F=u_{x}^{2}-2uu_{x}+uu_{xx}.
	\]
	The induced pseudospherical one-forms can be written as \cite{CastroSilva2015,CastroSilva2016}
	\begin{align*}
		\omega_{1} &= m~dx+F~dt, \\
		\omega_{2} &= (\mu m\pm2\sqrt{1+\mu^{2}})dx+\mu F~dt, \\
		\omega_{3} &= (\pm\sqrt{1+\mu^{2}}m+2\mu)dx\pm\sqrt{1+\mu^{2}}F~dt,
	\end{align*}
	where $\mu\in\mathbb{R}$ is a geometric parameter.
	
	A direct computation from the preceding expressions gives
	\[
	\omega_{1}\wedge\omega_{2}=\mp2\sqrt{1+\mu^{2}}F~dx\wedge dt.
	\]
	Therefore, the coframe non-degeneracy condition is equivalent to
	\[
	F(x,t) \neq 0.
	\]
	This observation will be crucial in the proof below. 
	Indeed, to ensure that the pseudospherical metric induced by a solution of the DP equation 
	is well defined on a given region, one has to establish the non-vanishing of the function
	\[
	F=u_x^2-2u u_x+u u_{xx}
	\]
	on that region.
	\subsection{The Induced First Fundamental Form and Its Metric Components}
	The first fundamental form defined by the coframe $\omega_{1},\omega_{2}$ is
	\[
	g=\omega_{1}^{2}+\omega_{2}^{2}.
	\]
	In the coordinates \((x,t)\), we write the first fundamental form as
	\[
	g=g_{11}\,dx^{2}+2g_{12}\,dx\,dt+g_{22}\,dt^{2}.
	\]
	Substituting the expressions for \(\omega_1\) and \(\omega_2\) into 
	\(g=\omega_1^2+\omega_2^2\), we obtain
	\begin{align*}
		g_{11}
		&=(1+\mu^{2})m^{2}+4(1+\mu^{2})
		\pm 4\mu\sqrt{1+\mu^{2}}\,m,\\
		g_{12}
		&=\bigl[(1+\mu^{2})m\pm2\mu\sqrt{1+\mu^{2}}\bigr]F,\\
		g_{22}
		&=(1+\mu^{2})F^{2}.
	\end{align*}
	Here the choice of sign is the same as that in the definition of \(\omega_2\).
	Equivalently, using
	\[
	m=u-u_{xx},\qquad F=u_x^{2}-2u u_x+u u_{xx},
	\]
	we may write
	\begin{align*}
		g_{11}
		&=(1+\mu^{2})(u-u_{xx})^{2}
		+4(1+\mu^{2})
		\pm4\mu\sqrt{1+\mu^{2}}\,(u-u_{xx}),\\
		g_{12}
		&=\bigl[(1+\mu^{2})(u-u_{xx})
		\pm2\mu\sqrt{1+\mu^{2}}\bigr]
		\bigl(u_x^{2}-2u u_x+u u_{xx}\bigr),\\
		g_{22}
		&=(1+\mu^{2})
		\bigl(u_x^{2}-2u u_x+u u_{xx}\bigr)^{2}.
	\end{align*}
    It follows that the metric component \(g_{22}\) is completely controlled by the square of the key function \(F\). 
    Consequently, in the proof below, once we prove that \(F(\gamma(t))\to+\infty\) along the characteristic curve \(\gamma(t)\) under consideration, it follows immediately that
    \[
     g_{22}(\gamma(t))\to+\infty.
    \]
    Moreover, along the characteristic constructed in the proof, one has \(m(\gamma(t))=0\). Hence,
    \[
    g_{12}(\gamma(t))
    =
    \pm 2\mu\sqrt{1+\mu^2}\,F(\gamma(t)),
    \]
    and therefore, when \(\mu\neq0\),
    \[
    |g_{12}(\gamma(t))|\to+\infty.
    \]
	
	\section{The Metric Blow-Up Theorem}
	\label{sec:theorem_statement}
	This section states the main result of the paper. 
	The theorem shows that, under suitable one-sided sign conditions on the initial momentum, 
	finite-time wave breaking for a solution of the DP equation leads to blow-up of certain components 
	of the induced pseudospherical metric.
	
	\begin{theorem}[Finite-Time Blow-Up of the Induced Metric]
		\label{thm:blowup}
		Let \(u_{0}\in H^{4}(\mathbb{R})\), and set
		\[
		m_{0}=u_{0}-u_{0}^{\prime\prime}.
		\]
		Let \(u\) be the corresponding strong solution of the DP equation on its maximal interval of existence \([0,T_{\max})\), written in the nonlocal form
		\[
		u_t+u u_x=-\frac{3}{2}\partial_x(1-\partial_x^2)^{-1}(u^2),
		\qquad x\in\mathbb{R},\quad 0<t<T_{\max},
		\]
		with initial condition
		\[
		u(x,0)=u_0(x).
		\]
		Set
		\[
		\mathcal{S}=\mathbb{R}\times[0,T_{\max}).
		\]
		
		For a fixed parameter \(\mu\in\mathbb{R}\) and a fixed choice of sign, define
		\[
		F=u_x^2-2u u_x+u u_{xx},
		\]
		and introduce the one-forms
		\[
		\omega_{1}=(u-u_{xx})\,dx+F\,dt,
		\]
		\[
		\omega_{2}=
		\bigl(\mu(u-u_{xx})\pm2\sqrt{1+\mu^{2}}\bigr)\,dx+\mu F\,dt.
		\]
		On any subregion of \(\mathcal{S}\) where \(F\neq0\), the one-forms
		\(\omega_1\) and \(\omega_2\) form a local coframe and define the induced first fundamental form
		\[
		g=\omega_1^2+\omega_2^2.
		\]
		In the coordinates \((x,t)\), we write
		\[
		g=g_{11}\,dx^{2}+2g_{12}\,dx\,dt+g_{22}\,dt^{2}.
		\]
		
		Assume that there exists \(x_0\in\mathbb{R}\) such that
		\[
		m_0(x)\ge0,\qquad x\le x_0,
		\]
		and
		\[
		m_0(x)\le0,\qquad x\ge x_0.
		\]
		Assume moreover that
		\[
		m_0\not\equiv0 \quad \text{on } (-\infty,x_0),
		\qquad
		m_0\not\equiv0 \quad \text{on } (x_0,\infty).
		\]
		
		Then \(T_{\max}<\infty\). Set
		\[
		T_0=T_{\max}.
		\]
		Let \(q:[0,T_0)\to\mathbb{R}\) be the characteristic curve satisfying
		\[
		q'(t)=u(q(t),t),\qquad q(0)=x_0.
		\]
		With
		\[
		\gamma(t)=(q(t),t),
		\]
		one has
		\[
		\gamma([0,T_0))\subset\mathcal{S}.
		\]
		Moreover,
		\[
		F(\gamma(t))>0,\qquad 0\le t<T_0,
		\]
		so the induced coframe is non-degenerate along \(\gamma\). Furthermore,
		\[
		\lim_{t\to T_0^-} g_{22}(\gamma(t))=+\infty.
		\]
		If \(\mu\neq0\), then
		\[
		\lim_{t\to T_0^-}|g_{12}(\gamma(t))|=+\infty.
		\]
	\end{theorem}
	
\noindent\textbf{Idea of the proof.} 
The proof is based on the following key observations. 
Let \(q(t)\) be the characteristic starting from \(x_0\), and set
\[
\gamma(t)=(q(t),t).
\]
We shall use the shorthand \(h(\gamma(t))=h(q(t),t)\) for functions \(h=h(x,t)\).
The one-sided sign conditions on the initial momentum, together with the momentum-transport formula, imply that
\[
m(\gamma(t))=0
\]
along this characteristic. Hence,
\[
u_{xx}(\gamma(t))=u(\gamma(t)).
\]
Consequently, the function
\[
F=u_x^2-2u u_x+u u_{xx}
\]
reduces along \(\gamma(t)\) to
\[
F(\gamma(t))
=
\bigl(u(\gamma(t))-u_x(\gamma(t))\bigr)^2.
\]
Using the Green's-function representation and integral estimates along the characteristic, 
one derives a Riccati-type differential inequality for a suitable auxiliary quantity \(Y(t)\), 
which yields
\[
Y(t)\to+\infty
\qquad \text{as } t\to T_0^-.
\]
Since the wave height \(u\) remains bounded before wave breaking and \(u_x(q(t),t)<0\) along the characteristic, it follows that
\[
u_x(q(t),t)\to-\infty
\qquad \text{as } t\to T_0^-.
\]
Therefore,
\[
u(q(t),t)-u_x(q(t),t)\to+\infty,
\]
and hence
\[
F(\gamma(t))\to+\infty.
\]
Since
\[
g_{22}(\gamma(t))=(1+\mu^2)F(\gamma(t))^2,
\]
we obtain
\[
g_{22}(\gamma(t))\to+\infty.
\]
Moreover, along the same characteristic one has \(m(\gamma(t))=0\), and hence
\[
g_{12}(\gamma(t))
=
\pm2\mu\sqrt{1+\mu^2}\,F(\gamma(t)).
\]
Thus, when \(\mu\neq0\),
\[
|g_{12}(\gamma(t))|\to+\infty.
\]
	
	\section{Proof of the Metric Blow-Up Theorem}
	\label{sec:proof}
	
	This section proves Theorem~\ref{thm:blowup}. 
	The proof is divided into four steps. 
	First, we construct the relevant characteristic curve and derive the consequences of the initial sign conditions. 
	Second, we establish differential inequalities along this characteristic and prove the persistence of the sign structure. 
	Third, we introduce a suitable auxiliary quantity and derive a Riccati-type inequality that yields finite-time blow-up. 
	Finally, we convert this analytic blow-up into blow-up of the corresponding components of the induced pseudospherical metric.
	
	\subsection*{Step 1: Characteristic Curve and Consequences of the Sign Conditions}
	
	Let \(T_{\max}\) be the maximal lifespan of the strong solution of the DP equation. 
	Define the characteristic curve,
	\[
	\gamma(t)=(q(t),t),
	\]
	where the spatial coordinate \(q(t)\) satisfies
	\[
	q'(t)=u(q(t),t),\qquad q(0)=x_{0}.
	\]
	For any function \(h=h(x,t)\), we write
	\[
	h(\gamma(t))=h(q(t),t).
	\]
	
	In order to use the evolution formula for the momentum along characteristics, let
	\(Q(t,\xi)\) denote the full characteristic flow generated by the velocity field \(u\):
	\[
	Q_{t}(t,\xi)=u(Q(t,\xi),t),\qquad Q(0,\xi)=\xi.
	\]
	Differentiating with respect to \(\xi\) gives
	\[
	\frac{d}{dt}Q_{\xi}(t,\xi)
	=
	u_{x}(Q(t,\xi),t)Q_{\xi}(t,\xi),
	\]
	and therefore
	\[
	Q_{\xi}(t,\xi)
	=
	\exp\left(\int_{0}^{t}u_{x}(Q(s,\xi),s)\,ds\right)>0.
	\]
	Thus, for each fixed \(t\), the map \(\xi\mapsto Q(t,\xi)\) is strictly increasing.
	
	Set
	\[
	m=u-u_{xx}.
	\]
	The momentum-transport form of the DP equation is
	\[
	m_{t}+um_{x}+3u_{x}m=0.
	\]
	Computing along the characteristic flow \(Q(t,\xi)\) yields
	\[
	\frac{d}{dt}m(Q(t,\xi),t)
	=
	-3u_{x}(Q(t,\xi),t)m(Q(t,\xi),t).
	\]
	On the other hand,
	\[
	\frac{d}{dt}Q_{\xi}(t,\xi)^{3}
	=
	3u_{x}(Q(t,\xi),t)Q_{\xi}(t,\xi)^{3}.
	\]
	Consequently,
	\[
	\frac{d}{dt}\left[m(Q(t,\xi),t)Q_{\xi}(t,\xi)^{3}\right]=0.
	\]
	We obtain the evolution formula for the momentum along characteristics:
	\[
	m(Q(t,\xi),t)Q_{\xi}(t,\xi)^{3}=m_{0}(\xi).
	\]
	
	Taking \(\xi=x_{0}\), and using the sign assumptions
	\[
	m_{0}(x)\ge 0,\qquad x\le x_{0},
	\]
	and
	\[
	m_{0}(x)\le 0,\qquad x\ge x_{0},
	\]
	we have, in particular,
	\[
	m_{0}(x_{0})=0.
	\]
	Hence
	\[
	m(Q(t,x_{0}),t)Q_{\xi}(t,x_{0})^{3}=0.
	\]
	Since \(Q_{\xi}(t,x_{0})>0\), it follows that
	\[
	m(Q(t,x_{0}),t)=0.
	\]
	Writing
	\[
	q(t)=Q(t,x_{0}),
	\]
	we therefore have, along the characteristic curve \(\gamma(t)=(q(t),t)\),
	\[
	m(\gamma(t))=0.
	\]
	That is,
	\[
	u(\gamma(t))-u_{xx}(\gamma(t))=0,
	\]
	and hence
	\[
	u_{xx}(\gamma(t))=u(\gamma(t)).
	\]
	
	Using \(m=u-u_{xx}\) and the Green's function
	\[
	G(x)=\frac{1}{2}e^{-|x|},
	\]
	we have
	\[
	u=G*m.
	\]
	Therefore \(u\) and \(u_x\) may be represented as
	\[
	u(x,t)
	=
	\frac{e^{-x}}{2}\int_{-\infty}^{x}e^{z}m(z,t)\,dz
	+
	\frac{e^{x}}{2}\int_{x}^{\infty}e^{-z}m(z,t)\,dz,
	\]
	and
	\[
	u_{x}(x,t)
	=
	-\frac{e^{-x}}{2}\int_{-\infty}^{x}e^{z}m(z,t)\,dz
	+
	\frac{e^{x}}{2}\int_{x}^{\infty}e^{-z}m(z,t)\,dz.
	\]
	It follows that
	\[
	u(x,t)+u_{x}(x,t)
	=
	e^{x}\int_{x}^{\infty}e^{-z}m(z,t)\,dz,
	\]
	and
	\[
	u(x,t)-u_{x}(x,t)
	=
	e^{-x}\int_{-\infty}^{x}e^{z}m(z,t)\,dz.
	\]
	
	Define two key quantities evolving along the characteristic curve:
	\[
	I(t):=(u+u_{x})(q(t),t)
	=
	e^{q(t)}\int_{q(t)}^{\infty}e^{-z}m(z,t)\,dz,
	\]
	and
	\[
	g(t):=(u-u_{x})(q(t),t)
	=
	e^{-q(t)}\int_{-\infty}^{q(t)}e^{z}m(z,t)\,dz.
	\]
	Here \(g(t)\) is an auxiliary function along the characteristic curve and should not be confused with the first fundamental form \(g\).
	
	Substituting \(t=0\) and using the initial sign conditions, we obtain
	\[
	I(0)
	=
	e^{x_{0}}\int_{x_{0}}^{\infty}e^{-z}m_{0}(z)\,dz
	<0,
	\]
	because
	\[
	m_{0}(z)\le 0,\qquad z\ge x_{0},
	\]
	and \(m_0\not\equiv0\) on \((x_0,\infty)\). Similarly,
	\[
	g(0)
	=
	e^{-x_{0}}\int_{-\infty}^{x_{0}}e^{z}m_{0}(z)\,dz
	>0,
	\]
	because
	\[
	m_{0}(z)\ge 0,\qquad z\le x_{0},
	\]
	and \(m_0\not\equiv0\) on \((-\infty,x_0)\). Thus, at the initial time,
	\[
	I(0)<0,\qquad g(0)>0.
	\]
	
	\subsection*{Step 2: Differential Inequalities and Preservation of the Sign Structure}
	
	Taking total derivatives of \(I(t)\) and \(g(t)\) with respect to time, and using the
	Green's-function representation to estimate the nonlocal term in the DP equation, we obtain
	the following two key differential inequalities:
	\[
	I'(t)\le \frac{1}{2}I(t)g(t),
	\]
	and
	\[
	g'(t)\ge -\frac{1}{2}I(t)g(t).
	\]
	
	Since the initial values satisfy
	\[
	I(0)<0,\qquad g(0)>0,
	\]
	these inequalities imply the persistence of the sign structure. Indeed, suppose that on some
	time interval one has
	\[
	I(t)<0,\qquad g(t)>0.
	\]
	Then
	\[
	I(t)g(t)<0.
	\]
	Consequently,
	\[
	I'(t)\le \frac{1}{2}I(t)g(t)<0,
	\]
	so \(I(t)\) is decreasing on this interval and hence
	\[
	I(t)\le I(0)<0.
	\]
	Similarly,
	\[
	g'(t)\ge -\frac{1}{2}I(t)g(t)>0,
	\]
	so \(g(t)\) is increasing on this interval and hence
	\[
	g(t)\ge g(0)>0.
	\]
	By a standard continuity argument, the sign structure persists throughout the interval of
	existence:
	\[
	I(t)<0,\qquad g(t)>0.
	\]
	
	Consequently, we obtain the sign of the spatial slope along the characteristic curve. By
	definition,
	\[
	I(t)-g(t)
	=
	(u+u_x)(q(t),t)-(u-u_x)(q(t),t)
	=
	2u_x(q(t),t).
	\]
	Since \(I(t)<0\) and \(g(t)>0\), it follows that
	\[
	2u_x(q(t),t)=I(t)-g(t)<0.
	\]
	Therefore
	\[
	u_x(q(t),t)<0.
	\]

	\subsection*{Step 3: A Riccati-Type Inequality and Finite-Time Blow-Up of the Auxiliary Quantity}
	
	To prove finite-time blow-up of the auxiliary quantity, define
	\[
	Y(t):=-I(t)g(t).
	\]
	Since \(I(t)<0\) and \(g(t)>0\), we have
	\[
	Y(t)>0.
	\]
	On the other hand,
	\[
	Y(t)
	=
	-\bigl[(u+u_x)(u-u_x)\bigr](q(t),t)
	=
	u_x(q(t),t)^2-u(q(t),t)^2.
	\]
	
	Differentiating \(Y(t)\) gives
	\[
	Y'(t)
	=
	-I'(t)g(t)-I(t)g'(t).
	\]
	Using the differential inequalities obtained in Step 2, and observing that
	\[
	-g(t)<0,\qquad -I(t)>0,
	\]
	we obtain
	\[
	Y'(t)
	\ge
	-\left(\frac{1}{2}I(t)g(t)\right)g(t)
	-
	I(t)\left(-\frac{1}{2}I(t)g(t)\right).
	\]
	After simplification,
	\[
	Y'(t)
	\ge
	-\frac{1}{2}I(t)g(t)\bigl[g(t)-I(t)\bigr].
	\]
	Since
	\[
	Y(t)=-I(t)g(t),
	\]
	it follows that
	\[
	Y'(t)
	\ge
	\frac{1}{2}Y(t)\bigl[g(t)-I(t)\bigr].
	\]
	Moreover,
	\[
	g(t)-I(t)
	=
	(u-u_x)(q(t),t)-(u+u_x)(q(t),t)
	=
	-2u_x(q(t),t).
	\]
	Hence
	\[
	Y'(t)
	\ge
	\frac{1}{2}Y(t)\bigl[-2u_x(q(t),t)\bigr].
	\]
	
	From Step 2 we already know that
	\[
	u_x(q(t),t)<0.
	\]
	Since
	\[
	Y(t)=u_x(q(t),t)^2-u(q(t),t)^2>0,
	\]
	we have
	\[
	u_x(q(t),t)^2
	=
	Y(t)+u(q(t),t)^2
	\ge
	Y(t).
	\]
	Combining this with \(u_x(q(t),t)<0\), we obtain
	\[
	-u_x(q(t),t)
	=
	\sqrt{u_x(q(t),t)^2}
	\ge
	\sqrt{Y(t)}.
	\]
	Substituting this into the preceding inequality yields the Riccati-type differential inequality
	\[
	Y'(t)
	\ge
	\frac{1}{2}Y(t)\bigl(2\sqrt{Y(t)}\bigr)
	=
	Y(t)^{3/2}.
	\]
	
	Since \(Y(t)>0\), we may separate variables:
	\[
	Y(t)^{-3/2}Y'(t)\ge 1.
	\]
	Integrating from \(0\) to \(t\), we get
	\[
	\int_{0}^{t}Y(s)^{-3/2}Y'(s)\,ds
	\ge
	\int_{0}^{t}1\,ds.
	\]
	Thus
	\[
	-2Y(t)^{-1/2}+2Y(0)^{-1/2}\ge t.
	\]
	Equivalently,
	\[
	\frac{2}{\sqrt{Y(t)}}
	\le
	\frac{2}{\sqrt{Y(0)}}-t.
	\]
	Since the left-hand side is positive, the right-hand side must remain positive as long as \(Y(t)\) remains finite. Therefore \(Y(t)\) cannot remain finite beyond the time
	\[
	\frac{2}{\sqrt{Y(0)}}.
	\]
	Consequently, there exists a finite blow-up time \(T_0\) satisfying
	\[
	T_0\le \frac{2}{\sqrt{Y(0)}}<\infty
	\]
	such that
	\[
	\lim_{t\to T_0^-}Y(t)=+\infty.
	\]
	
	\subsection*{Step 4: From Analytic Blow-Up to Metric Component Blow-Up}
	
	First, from
	\[
	Y(t)=u_x(q(t),t)^2-u(q(t),t)^2
	\]
	and
	\[
	Y(t)\to+\infty
	\qquad \text{as } t\to T_0^-,
	\]
	we see that, since the wave height \(u(x,t)\) remains bounded before wave breaking,
	\[
	u_x(q(t),t)^2\to+\infty.
	\]
	By the wave-breaking theory for the DP equation and the a priori bounds supplied by the
	corresponding conservation laws, \(u(x,t)\) remains bounded before breaking\cite{Liu2006,Liu2007,Yin2003,Yin2004}. Together with
	the fact already proved in Step 2 that
	\[
	u_x(q(t),t)<0,
	\]
	this gives
	\[
	u_x(q(t),t)\to-\infty
	\qquad \text{as } t\to T_0^-.
	\]
	
	We now examine the geometric factor. The key geometric quantity considered in the paper is
	\[
	F=u_x^2-2u u_x+u u_{xx}.
	\]
	Along the characteristic curve \(\gamma(t)=(q(t),t)\), Step 1 established that
	\[
	m(\gamma(t))=0.
	\]
	Since \(m=u-u_{xx}\), we have
	\[
	u_{xx}(\gamma(t))=u(\gamma(t)).
	\]
	Therefore
	\[
	F(\gamma(t))
	=
	u_x(\gamma(t))^2
	-
	2u(\gamma(t))u_x(\gamma(t))
	+
	u(\gamma(t))^2.
	\]
	That is,
	\[
	F(\gamma(t))
	=
	\bigl(u(\gamma(t))-u_x(\gamma(t))\bigr)^2.
	\]
	Define
	\[
	f(t):=\bigl(u(q(t),t)-u_x(q(t),t)\bigr)^2.
	\]
	Then
	\[
	F(\gamma(t))=f(t).
	\]
	Since \(u(q(t),t)\) is bounded and
	\[
	u_x(q(t),t)\to-\infty
	\qquad \text{as } t\to T_0^-,
	\]
	we obtain
	\[
	f(t)\to+\infty
	\qquad \text{as } t\to T_0^-.
	\]
	In particular,
	\[
	F(\gamma(t))=f(t)>0,\qquad 0\le t<T_0,
	\]
	so the induced coframe is non-degenerate along \(\gamma\) before the blow-up time.
	
	By the formula for the metric components, along the characteristic curve \(\gamma(t)\) we have
	\[
	g_{22}(\gamma(t))
	=
	(1+\mu^2)f(t)^2.
	\]
	Moreover, since \(m(\gamma(t))=0\), the mixed component reduces to
	\[
	g_{12}(\gamma(t))
	=
	\pm 2\mu\sqrt{1+\mu^2}\,f(t).
	\]
	Since
	\[
	f(t)\to+\infty
	\qquad \text{as } t\to T_0^-,
	\]
	we conclude that
	\[
	\lim_{t\to T_0^-}g_{22}(\gamma(t))=+\infty.
	\]
	Moreover, when \(\mu\neq0\),
	\[
	\lim_{t\to T_0^-}|g_{12}(\gamma(t))|=+\infty.
	\]
	
	Thus, along the characteristic curve whose spatial component starts from \(x_0\), the metric
	component \(g_{22}\) diverges to \(+\infty\) at the finite-time limit. When \(\mu\neq0\), the
	absolute value of the mixed component \(g_{12}\) also diverges to \(+\infty\). This proves
	Theorem~\ref{thm:blowup}.
	\hfill\(\Box\)
	
	\section{Geometric Interpretation and Local Isometric Immersions}
	\label{sec:geometric_interpretation}
	
	This section further explains the geometric meaning of the result, especially its relation to the first fundamental form, coframe non-degeneracy, and local isometric immersions. Related geometric breakdown and immersion phenomena have also been studied for other integrable shallow water equations: in particular, Freire analyzed the breakdown of pseudospherical surfaces determined by the CH equation, while Freire and Tito studied a Novikov equation describing pseudospherical surfaces, its pseudo-potentials, and local isometric immersions \cite{Freire2024,Freire2022}.
	
	\subsection{Metric Blow-Up Is Not Curvature Blow-Up}
	
	By the theory of pseudospherical equations, if a triple of one-forms 
	\(\omega_1,\omega_2,\omega_3\) satisfies the structure equations and 
	\(\omega_1\wedge\omega_2\neq0\), then
	\[
	g=\omega_1^2+\omega_2^2
	\]
	defines a pseudospherical metric of Gaussian curvature \(K=-1\). Therefore, the blow-up 
	proved in Theorem~\ref{thm:blowup} is not a blow-up of the Gaussian curvature itself. 
	Rather, what becomes unbounded are certain coefficients of the first fundamental form 
	in the coordinates \((x,t)\).
	
	Along the characteristic curve \(\gamma(t)\), one has
	\[
	g_{22}(\gamma(t))\to+\infty
	\qquad \text{as } t\to T_0^-.
	\]
	Thus, in the induced coordinate representation, the coefficient of \(dt^2\), equivalently 
	the squared length of the coordinate vector \(\partial_t\), grows without bound along 
	\(\gamma(t)\). The geometric singularity is therefore not a curvature singularity, since 
	the curvature remains normalized to \(K=-1\) on the non-degenerate precritical region. 
	It is instead a blow-up of metric components of the first fundamental form at the 
	finite-time boundary.
	
	\subsection{Coframe Non-Degeneracy and Metric Component Blow-Up}
	
	The coframe non-degeneracy condition is equivalent to
	\[
	F\neq0.
	\]
	Along the characteristic curve \(\gamma(t)\) constructed in the proof, we have
	\[
	F(\gamma(t))=f(t)>0,\qquad 0\le t<T_0.
	\]
	This shows that, before the blow-up time, the induced coframe does not degenerate along 
	the relevant characteristic. In particular, the metric is well defined there. As
	\[
	t\to T_0^-,
	\]
	we have
	\[
	F(\gamma(t))\to+\infty,
	\]
	and consequently
	\[
	g_{22}(\gamma(t))\to+\infty.
	\]
	Thus, the geometric singularity in this problem is not a degeneracy-type singularity caused by
	\[
	\omega_1\wedge\omega_2=0.
	\]
	Rather, it is a blow-up-type singularity caused by the unbounded growth of metric components.
	
	\subsection{From Wave Breaking to Metric Blow-Up}
	
	The proof establishes a precise correspondence between wave breaking for the DP equation 
	and blow-up of the induced pseudospherical metric components. Along the characteristic 
	curve \(\gamma(t)\), the mechanism is
	\[
	u_x(q(t),t)\to-\infty
	\quad \Longrightarrow \quad
	F(\gamma(t))\to+\infty
	\quad \Longrightarrow \quad
	g_{22}(\gamma(t))\to+\infty.
	\]
	Moreover, when \(\mu\neq0\), the identity
	\[
	g_{12}(\gamma(t))
	=
	\pm 2\mu\sqrt{1+\mu^2}\,F(\gamma(t))
	\]
	shows that the mixed component also blows up in absolute value:
	\[
	|g_{12}(\gamma(t))|\to+\infty.
	\]
	In this sense, the analytic singularity of the DP solution is transmitted to the first 
	fundamental form of the induced pseudospherical metric.
	
	\subsection{Consequences for Local Isometric Immersions}
	
	On any local region where the coframe is non-degenerate, the induced first fundamental form 
	may be interpreted as the intrinsic metric of a pseudospherical surface. Where a classical 
	local isometric immersion realizing this metric exists, its first fundamental form is precisely
	\[
	g=\omega_1^2+\omega_2^2.
	\]
	However, since
	\[
	g_{22}(\gamma(t))\to+\infty
	\qquad \text{as } t\to T_0^-,
	\]
	the corresponding local immersion structure cannot be extended through the critical-time 
	boundary in the classical sense while retaining finite coefficients of the first fundamental 
	form in the induced coordinates~\cite{CastroSilva2016,Freire2025}. In other words, at the level of the induced geometry, 
	wave breaking prevents a classical continuation of the associated local isometric immersion 
	structure with finite metric components along the characteristic curve.
	
	\section{Conclusion}
	\label{sec:conclusion}
	
	This paper has studied the blow-up of pseudospherical metrics induced by solutions of the Cauchy problem for the DP equation. 
	For initial momentum profiles satisfying suitable left--right sign conditions, we used the method of characteristics, the momentum-transport formula, the Green's-function representation, and Riccati-type differential inequalities to prove that certain components of the induced metric blow up in finite time. 
	More precisely, along the characteristic curve constructed in the proof, the metric component \(g_{22}\) diverges to \(+\infty\), and when \(\mu\neq0\), the mixed component \(g_{12}\) also blows up in absolute value. 
	The blow-up obtained here is not a curvature blow-up, since the Gaussian curvature remains normalized to \(K=-1\) on the non-degenerate precritical region; rather, it is a metric-component blow-up occurring in the induced first fundamental form. 
	This shows that wave breaking of solutions of the DP equation manifests itself not only as unbounded growth of the spatial slope in the PDE sense, but also through the induced one-form structure as blow-up of certain components of the first fundamental form of a pseudospherical metric. 
	Thus, this paper establishes a rigorous connection between finite-time wave breaking for the DP equation and blow-up of its induced pseudospherical metric.
	
	
	\section*{Acknowledgments}
	
	This work was supported by Research Project of the Shaanxi Institute of Basic Science (NO.25JSY050).


\end{document}